\documentstyle[11pt,newpasp,twoside,epsf,psfig]{article}

\def\edcomment#1{\iffalse\marginpar{\raggedright\sl#1\/}\else\relax\fi}
\reversemarginpar
\newcommand{\HI}{\protect\normalsize H\thinspace\protect\footnotesize
I\protect\normalsize} 
\newcommand{\kms}{\,km\,s$^{-1}$}
\newcommand{\etal}{{et~al.\, }}

\begin{document}

\title{A Search of the Zone of Avoidance in Scorpius}

\author{Anthony P. Fairall}
\affil{Department of Astronomy, University of Cape Town, Rondebosch 7700, South Africa}

\author{Ren\'ee C. Kraan-Korteweg}
\affil{Depto.\ de Astronom\' \i a, Universidad de Guanajuato, 
Apdo.~Postal 144, Guanajuato GTO 36000, Mexico}

\begin{abstract} An optical search of the Scorpius region -- close to
the Galactic bulge -- has revealed some 1400 partially-obscured
galaxies. Redshifts have been obtained for nearly a hundred of these
objects. Preliminary indications of large-scale structures are
reported.
\end{abstract} 

\section{Introduction}

As outlined in the introductory paper to this session 
(Kraan-Korteweg, these proceedings), deep optical galaxy searches 
in the Zone of Avoidance (ZOA) narrow the band of obscuration 
down to on average 5 or 6 degrees (see Fig.~4 in Kraan-Korteweg,
these proceedings). Moreover, optical searches have the advantage over both
\HI\ and infrared surveys that they can uncover clusters rich in E/S0
galaxies which generally trace the mass density peaks in the
Universe.  The nearby rich cluster ACO 3627, believed to lie at the
centre of the Great Attractor, is the most conspicuous case
(Kraan-Korteweg \etal 1996, Woudt et~al., these proceedings).

This conference has brought together all the researchers primarily
responsible for systematic optical searches -- Weinberger,
Kraan-Korteweg and Woudt, Saito and Wakamatsu. Thanks to their
industry, all but one small sector of the Milky Way had been surveyed
by 1998. The only remaining sector was the Scorpius region, between
Galactic longitudes of $330\deg \la \ell \la 350\deg$, outlined in
Fig.~1 (thick contour) in a distribution of galaxies larger than $D
\ge 1\farcm3$ centered on the southern Milky Way. The previous search
areas are indicated as well (for details see Kraan-Korteweg, these
proceedings). The Scorpius region was surveyed by the first author
(Fairall) in late 1998, whilst on sabbatical leave from the University
of Cape Town, based with Kraan Korteweg at the Department of Astronomy
of the University of Guanajuato. This was particularly opportune
because it allowed the use of the same viewing machine employed in the
earlier searches by Kraan-Korteweg (2000), and those carried out under
her direction by Woudt (1998) and Salem (Salem \& Kraan-Korteweg, in
prep.).

\begin{figure}
\plotfiddle{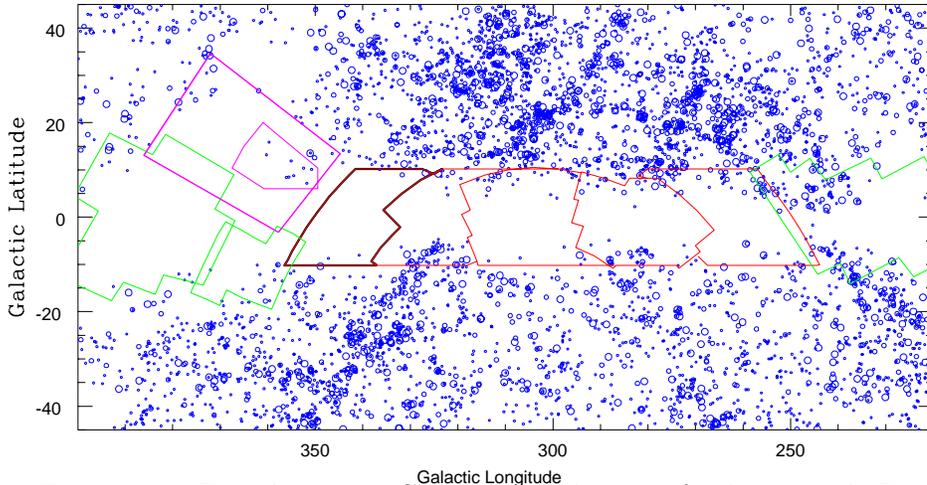}{6.5cm}{-90}{50}{50}{-200}{250}
\caption{Distribution in Galactic coordinates of galaxies with $D \ge
1\farcm3$. The ZOA (running horizontally through the centre of the
diagram) shows an absence of galaxies.  Various researchers have
conducted deep optical searches in the ZOA (marked areas). This
paper reports on the Scorpius search (thick contour) centred on $\ell
= 345\deg$.  Note the almost complete absence of Lauberts galaxies in
the region.}
\end{figure}

\section{The Search}

Sixteen fields of the SERC IIIa-J Sky Survey (226-228, 274-280 and
330-335) were searched within Galactic latitudes of $|b| < 10\deg$.
Small portions of Fields 068 and 181 -- omitted in earlier
neighbouring surveys -- were also searched. The viewing machine with
which the plates were examined displays a 50-times magnified
segment. Adjacent strips of the width of 4\,mm (on the plate) were
systematically searched for galaxies, offset by 3.5\,mm to allow some
overlap.  Working close to the Galactic bulge, the images reveal up to
several hundred stars on the screen at any one time. Amongst these,
the eye must discern the occasional fuzzy image of a galaxy.  It seems
a curiously old-fashioned technique to employ in this computer age,
but past experience has shown that computer algorithms, searching for
galaxies, are too easily confused by the abundance of overlapping
stellar images (see Kraan-Korteweg \& Lahav 2000). The eye is still
the most efficient means of reliably identifying partially obscured
galaxies.

As with the previous searches of Kraan-Korteweg and collaborators, a
diameter limit of 12\,arcsec (10\,mm on the screen) was
employed. However, diameter limited catalogues strongly favour edge-on
spirals.  Therefore elliptical and S0 galaxies with diameters somewhat
below the $12\arcsec$ limit were also recorded. Working at such low
Galactic latitude, planetary and reflection nebulae often mimic the
appearances of distant galaxies; these too were registered.

\begin{figure} [p]
\hfil \psfig {figure=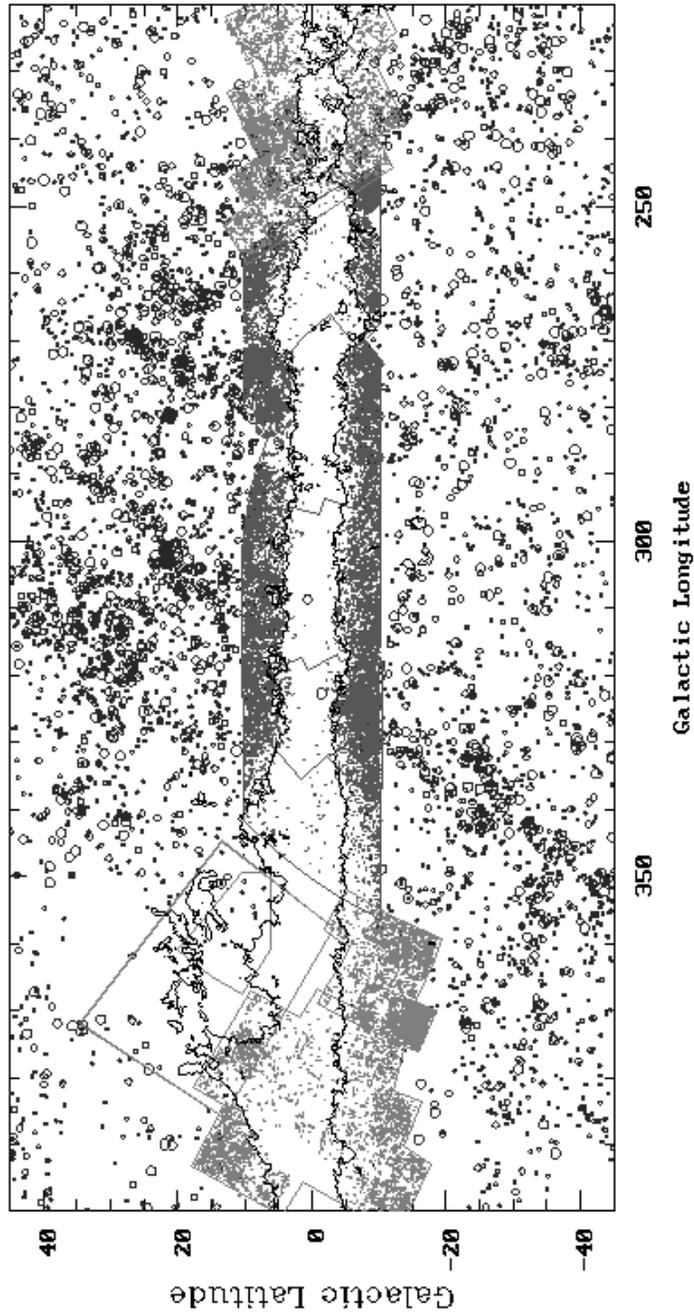,angle=90,width=10cm} \hfil
\caption{The same as Fig.~1, but now including galaxies with $D \ge
12\arcsec$ found in the optical searches. The ZOA has been narrowed 
to about a third of what it was previously. The extinction contour 
of  $A_B = 3\fm0$ (according to the DIRBE dust maps; Schlegel \etal 1998)
is also shown.}
\end{figure}

In all, about 1900 potential galaxies were listed. This
number includes objects counted twice in regions of overlapping fields
-- and also includes galaxy candidates on the borders of the previously surveyed
neighbouring Great Attractor region which will be used for consistency
check of the parameters of the uncovered galaxies. After removing
duplicates, and rejecting marginal objects, the number drops to 1422
galaxies, of which just below a thousand meet the diameter limit of $D
\ge 12\arcsec$. These are displayed in Fig.~2, together with galaxies
larger than $D \ge 12\arcsec$ identified in other optical galaxy
searches as well as the previously known galaxies with $D \ge 1\farcm3$.

As can be seen in Fig.~2, the yield in the Scorpius region is quite
modest, substantially below the average recorded in the adjacent
surveys further west (right in the diagram). The reason is obvious,
however. The obscuration of the Milky Way broadens considerably in the
region of the Galactic bulge (see the extinction contour $A_B = 3\fm0$ in
Fig.~2), particularly on the northern side of the Galactic plane.  The
reason for the drop-off on the southern side is not obvious, but a
$1\deg$ wide overlap with the neighbouring survey suggests it to be
real. It presumably reflects the presence of nearby void.

Preliminary statistics indicate:

\begin{itemize}
\item{Only 19 of the recorded galaxies have $D \ge 1\farcm0$, the
Lauberts limit.  Only 11 have $D \ge 1\farcm3$, the completeness limit
of the Lauberts catalogue.}

\item{104 galaxies have been found in extinction layers with $A_B >
3\fm0$. These most likely are heavily contaminated by Galactic objects.}

\item{Of the remainder, 977 have $D \ge 0\farcm2$, 13 have $D \ge
1\farcm0$, and 6 have $D \ge 1\farcm3$.}
\end{itemize}

\begin{figure}[t]
\hfil \psfig{figure=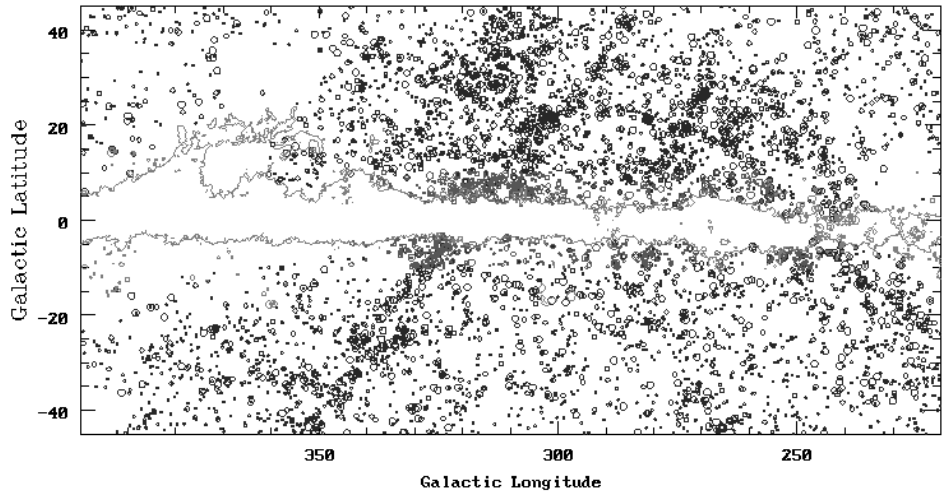,width=13cm} \hfil
\caption{Galaxies corrected for extinction with $D^o \ge 1\farcm3$
and $A_B \ge 3\fm0$, the completeness limit of deep optical searches. 
Only a reduced ZOA still obscures the view.}
\end{figure}

If corrected for extinction -- applying the Cameron (1990) laws --
these numbers would increase to 1294, 55 and 24 respectively.  The
last number allows us to extend the Lauberts galaxies plotted
originally in Fig.~1 to reflect the corrected sky distribution for
this region, except in the now narrower Zone of Avoidance, delimited
by $A_B = 3\fm0$. Figure 3 shows the overall picture.

\section{Redshifts}

One hundred of the brightest objects in the survey (including the
Field 068 and 181 annexures) have been observed spectroscopically in
June 1999 with the 1.9\,m reflector, UNIT spectrograph and CCD
detector at the South African Astronomical Observatory at
Sutherland. These allow preliminary plots on the distribution of the
recorded galaxies in redshift space. Figure 5 indicates the coverage
within our survey area. Figure 6 shows the same plot divided into
redshift slices of 2000~\kms\ thickness.

\begin{figure}[p]
\plotfiddle{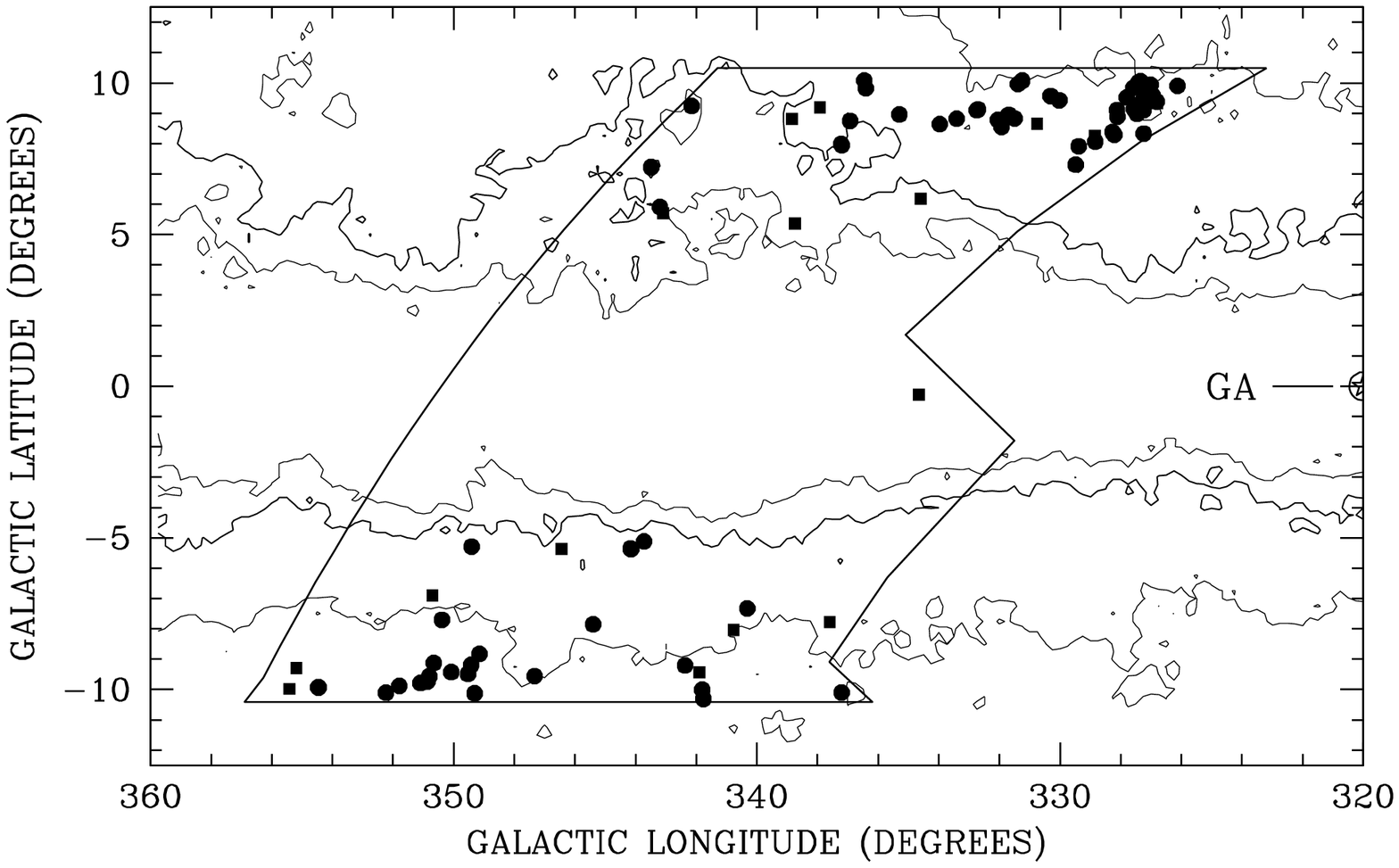}{6.8cm}{0}{55}{55}{-160}{-110}
\caption{Galaxies observed spectroscopically within the Scorpius survey
area. Filled squares indicate previously observed galaxies (from the
NED database). The filled circles indicate the new observations. The 
extinction contours indicate $A_B = 1\fm0; 3\fm0$ and $5\fm0$ respectively.}

\plotfiddle{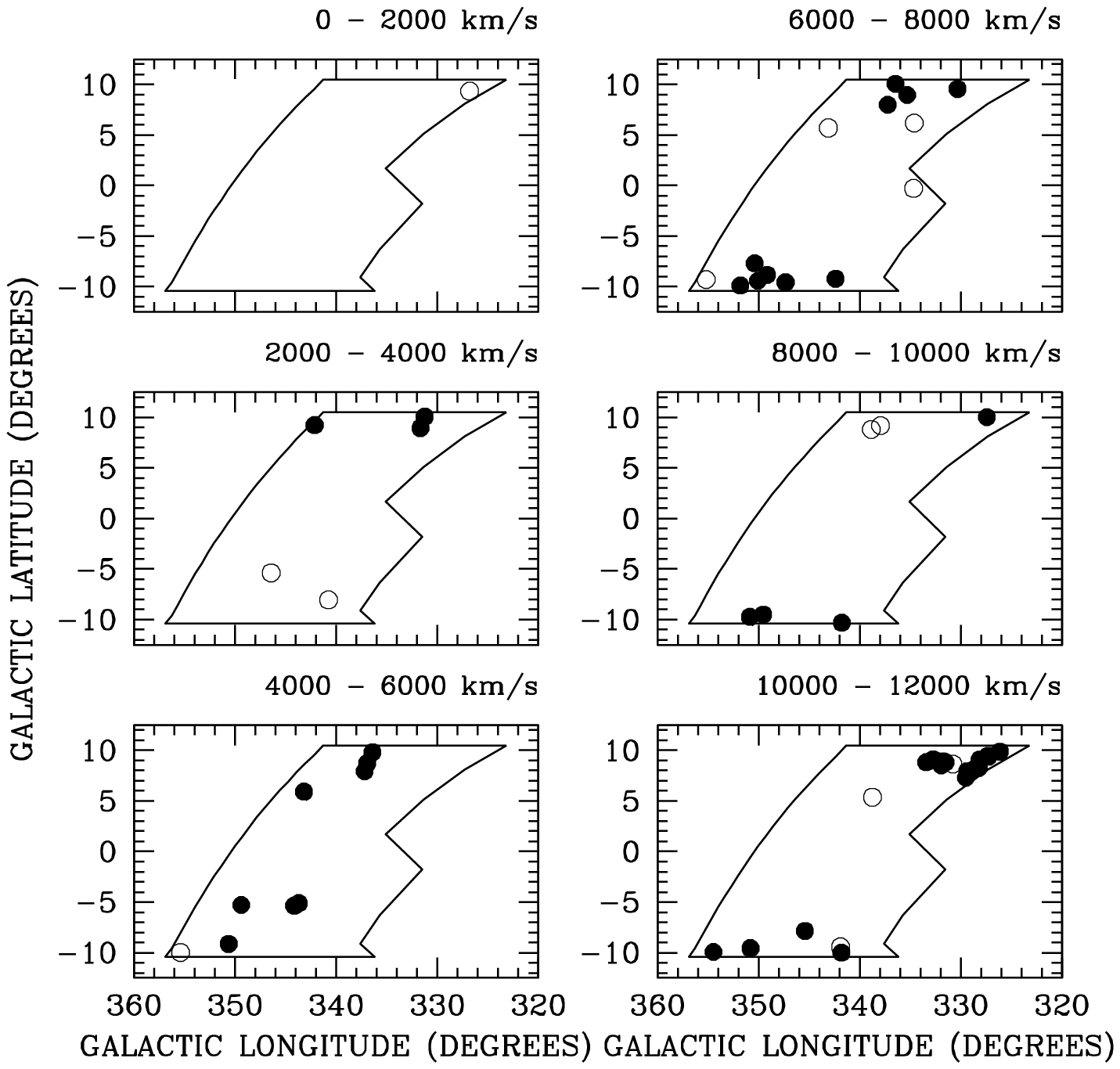}{10cm}{0}{75}{75}{-252}{-180}
\caption{As in Fig.~4, but shown as slices of increasing redshift in
increments of 2000~\kms.}
\end{figure}

As seen in Fig.~5, galaxies lie mainly in the 4000 -- 8000 or 10000
-- 12000 \kms\ redshift interval. The high redshift peak (at
11000~\kms) coincides with the highest surface density of galaxies
found in the survey (at $\ell = 330\deg$, $b = 8\deg$) and seems to
indicate a cluster complex, possibly a supercluster, in that
region. The lower redshift range is suggestive of a possible
overdensity around 6000~\kms. This is consistent with preliminary data
from the full sensitivity blind \HI\ survey of the southern ZOA ($|b|
< 5\deg$) performed at the Parkes 64\,m telescope (see Staveley-Smith
et~al., these proceedings) which also finds a concentration around
6000 -- 8000~\kms in that region. None of the optical observations
match the individual radio sources but the latter are detected at
lower latitudes, confirming that the optical plus \HI\ observations 
are complementary methods in tracing the galaxy distribution behind
the Milky Way.

A number of interesting objects have been uncovered by the survey. One
is an apparent cluster 1720$-$45 ($344\deg, -5\fdg3$) seen through the
bulge of the Milky Way. Only three galaxies, those with the highest
central surface brightnesses are optically detectable: one is a
marginal Seyfert galaxy, the other two are E/S0. Two of these were
spotted in the survey, the third was known previously from its
infrared emission (but was re-observed here).  Their redshifts are
5840, 5610 and 5633~\kms\ respectively, confirming the cluster nature.

The ESO (Lauberts) galaxies in the region are mainly spirals (seen
flat-on or close to flat-on), but which had no published
redshifts. ESO~274-G019 has 3389~\kms, ESO~330-059 has 6538~\kms and
ESO~330-061 5681~\kms.

Field 330 also revealed two curious objects, which look exactly like
elliptical galaxies but appear to stand in front dense dust clouds or were
seen through them. There the representative star count
-- normally 35 to 40 -- had dropped to 7. One of these objects has
a redshift of 3932~\kms, so clearly is extragalactic.

Follow-up work on these objects and at least one further season of
spectroscopic observations will be necessary to expand the sample
and discuss the galaxy distribution in redshift space in more
detail. We are also preparing the Scorpius galaxy data for publication
in a catalogue similar to that of Kraan-Korteweg (2000).

\acknowledgements
This research has made use of the NASA/IPAC Extragalactic Database (NED)
which is operated by the Jet Propulsion Laboratory, Caltech, under
contract with the National Aeronautics and Space Administration.
We are grateful to A. Schr\"oder for her preliminary plot of
\HI-detections in the Scorpius region.

\end{document}